\documentstyle[prl,aps,twocolumn,psfig]{revtex}
\begin{document}
\title{ON THE RELATION BETWEEN INTERBAND SCATTERING AND THE
"METALLIC PHASE" OF TWO DIMENSIONAL HOLES IN GaAs/AlGaAs }
\author{Yuval Yaish$^{a}$, Oleg Prus$^{a}$, Evgeny Buchstab$^{a}$, Shye Shapira$^{b}$, Gidi Ben Yoseph$^{a}$,
 Uri Sivan$^{a}$ and Ady Stern$^{c}$}
\address{$^{a}$Dep. of Physics and Solid State Institute, Technion-IIT, Haifa 32000,
Israel. \\
$^{b}$Cavendish Laboratory, Madingley Road, Cambridge CB3 OHE, UK. \\
$^{c}$Dep. of Condensed Matter Physics, Weizmann Institute of Science,
Rehovot 76100, Israel}

\maketitle

\begin{abstract}
The "metallic" regime of holes in GaAs/AlGaAs heterostructures corresponds
to densities where two splitted heavy hole bands exist at a zero magnetic field.
Using Landau fan diagrams and weak field magnetoresistance curves we extract
the carrier density in each band and the interband scattering rates. The
measured inelastic rates depend Arrheniusly on temperature with an
activation energy similar to that characterizing the longitudinal
resistance. The "metallic" characteristics, namely, the resistance
increase with temperature, is hence traced to the activation of inelastic
interband scattering. The data are used to extract the bands dispersion
relations as well as the two particle-hole excitation continua. It is then
argued that acoustic plasmon mediated Coulomb scattering might be
responsible for the Arrhenius dependence on temperature. The absence of
standard Coulomb scattering characterized by a power law dependence upon
temperature is pointed out.
\end{abstract}

\bigskip

\narrowtext

Non-interacting two dimensional electron gas (2DEG) 
is believed to be insulating in the sense that
 its resistance always diverges
as the temperature, $T$, approaches zero. This trend is opposite to that
characteristic of most 3D metals where for $T\rightarrow 0$ the resistance
becomes smaller and eventually saturates to a finite value. The insulating
nature of 2D systems has been observed in many experiments 
\cite{Bishop1,Pepper1} and was a generally accepted dogma until a few years ago,
when Kravchenko \textit{et} \textit{al}.  \cite{Kravchenko1} discovered that
in silicon based high mobility 2DEG at high enough carrier densities, the
resistance decreases as $T\rightarrow 0$ ("metallic" behavior). The same
samples, at lower densities, displayed insulating characteristics and the
crossover from positive to negative variation of the resistance with
temperature was soon identified as a novel metal-insulator transition.
Since the discovery of Kravchenko \textit{et al}.  \cite{Kravchenko1},
qualitatively similar dependencies of the resistance upon temperature were
observed in two dimensional hole gas (2DHG) in GaAs/AlGaAs heterostructures%
  \cite{Hanin1,Simmons1}, SiGe quantum wells  \cite{Koleridge1,Lam1}, InAs
quantum wells  \cite{Shayegan1} as well as other silicon samples  \cite{Popovic1}.

While the nature of the insulating phase may be roughly understood, the
physics of the metallic phase remains obscure.
A mere reduction of the resistance with temperature does not however
necessarily
imply delocalization. Within a Drude-like framework it may result from a
temperature dependent carrier density, or as suggested
by Altshuler and Maslov  \cite{Altshuler1}, by a temperature dependent
scattering time.

Here we provide experimental evidence that indeed suppression of scattering
with decreasing $T$ is responsible for the metallic characteristics of 2DHG
in GaAs/AlGaAs heterostructures. The scattering mechanism, interband Coulomb
scattering between the two splitted heavy hole bands, is very different from the one
proposed in ref. \cite{Altshuler1}. The correlation between
the existence of two conducting bands and "metallic" behavior was 
put forward by Pudalov  \cite{Pudalov1}. It has recently been convincingly
substantiated by Papadakis \textit{et} \textit{al}  \cite{Shayegan2}. We find
the same correlation but go beyond that and prove that the characteristic
dependence upon temperature, $\rho _{xx}(T)=\rho (0)+\alpha \exp (-T_{0}/T)$
($T_{0}$ and $\alpha $ are some constants) follows from a similar dependence
of the interband inelastic scattering rates upon temperature. We extract the
bands' structures and their particle-hole excitation continua from the
measurements. We then use the latter to show that the Arrhenius temperature
dependence might result from activation of plasmon mediated interaction,
similarly to plasmon enhanced Coulomb drag between coupled layers%
  \cite{Flensberg1}.

We believe that the same considerations may apply to the other "metallic"
2D systems since band degeneracy is generally lifted there due to spin-orbit
coupling and the lack of inversion symmetry at the interface where the 2D
layer resides.

The splitting between the two heavy hole bands in GaAs/AlGaAs heterostructures 
due to spin orbit coupling and lack of inversion symmetry
have been extensively
studied both theoretically  \cite{Broido1,Ekenberg1} and experimentally  \cite
{Stormer1,Shayegan3}. The two bands are approximately degenerate up to a certain energy
where they split and acquire very different effective masses (see inset to
fig. 4). Thus, above a threshold density, current is carried by two bands as
reflected in slope variation in the corresponding Landau fan diagram  \cite
{Stormer1} or in the appearance of a second frequency in Shubnikov de Haas
measurements  \cite{Shayegan3}. Using either method, the hole densities in the
lighter band, $p_{l\text{ }}$, and heavier band, $p_{h}$, can be extracted  \cite
{Shayegan3} and fig. 1 depicts them as a function of the total density, $%
p_{total}$. For a total density below $\simeq 1.7\times 10^{11}cm^{-2}$, the
bands are degenerate. For higher densities they split and for a total
density above $\simeq 2.8\times 10^{11}cm^{-2}$ , practically all additional
carriers go to the heavier and less mobile band. The inset to fig. 1 depicts
a characteristic Shubnikov de Haas curve and demonstrates the existence of
two sets of oscillations corresponding to the two bands. The data were taken
using a high mobility ($\simeq 500,000$ $cm^{2}/V\cdot s$ at $100$ $mK$)
2DHG confined to a $GaAs/Al_{0.8}Ga_{0.2}As$ interface in the 100 plane.
The sample had a 2DEG front gate and silicon doped backgate, 40nm and 
300nm from the 2DHG, respectively.

The simultaneous transport of two types of carriers with different
mobilities and densities is also manifested in our system by a weak field
classical positive magnetoresistance  \cite{Ashcroft1}. A set of resistance
curves for $p_{total}=4.25\times 10^{-11}cm^{-2}$, $p_{l}=1.52\times
10^{-11}cm^{-2}$, $p_{h}=2.73\times 10^{-11}cm^{-2}$ at different
temperatures is depicted in the inset to fig. 2. Measurements as a function
of density show that the positive magnetoresistance at weak fields appears
when the total density is beyond the split off density, namely, when the two
bands have different masses and mobilities. It then grows larger with
density, as the bands deviate further. The Lorentzian shaped
magnetoresistance expected from two band transport is obtained by
subtracting the weak parabolic negative magnetoresistance (attributed to
Coulomb interactions  \cite{Paalanen1}) from the full magnetoresistance curve.
It is depicted in
fig. 2. Note the excellent agreement with the predicted shape, eqs. 2 below.
Below $\simeq 0.6K$, the resistance is practically independent of $T$ while
for temperatures above $\simeq 2K$, the Lorentzian is hardly visible. The
suppression of the classical two band magnetoresistance results from
interband scattering. At low temperatures this scattering is mainly elastic.
As the temperature is increased, inelastic scattering commences, the drift
velocities of carriers in the two bands gradually approach each other, and
the magnetoresistance is consequently diminished.

The data presented in figs. 1 and 2 can be used to extract the interband
scattering rates. To that end, the standard two band transport formulae  \cite
{Ashcroft1} should be generalized to include interband scattering. The
starting point is coupled Drude equations for transport in the two bands
which straightforwardly give the resistance tensor

\begin{equation}
\left[ 
\begin{array}{l}
\varepsilon _{x} \\ 
\varepsilon _{y} \\ 
\varepsilon _{x} \\ 
\varepsilon _{y}
\end{array}
\right] = \left[ 
\begin{array}{cccc}
S_{l} & -HR_{l} & -Q & 0 \\ 
HR_{l} & S_{l} & 0 & -Q \\ 
-Q & 0 & S_{h} & -HR_{h} \\ 
0 & -Q & HR_{h} & S_{h}
\end{array}
\right] \left[ 
\begin{array}{l}
J_{lx} \\ 
J_{ly} \\ 
J_{hx} \\ 
J_{hy}
\end{array}
\right]  \label{rtensor}
\end{equation}

\smallskip \noindent Here, $\varepsilon $ and $H$ are the electric and
magnetic fields, respectively, $R_{i}=1/en_{i}$ $(i=l,h)$ is the Hall
coefficient of the $i$-th band, $n_{i}$ is the hole density in that band and 
$J_{i}$ is the current density there. The elements $S_{l}$, $S_{h}$, and $Q$
can be expressed in terms of conductances, $S_{l}=\sigma _{ll}^{-1}+\sigma
_{lh}^{-1};\ S_{h}=\sigma _{hh}^{-1}+\sigma _{hl}^{-1};Q=\lambda \sigma
_{lh}^{-1}=\lambda ^{-1}\sigma _{hl}^{-1}$ where $\lambda $ is a function of
the velocities and densities. The diagonal conductances, $\sigma _{ii}$,
pertain to scattering (elastic and inelastic) within each band while $\sigma
_{ij}$ accounts for interband scattering. The latter processes may include
carrier transfer from one band to the other as well as drag-like processes
where one particle from one band scatters off a particle in the other band
and both carriers maintain their bands. The function $\lambda $ depends on
the nature of the dominant interband scattering mechanism.

Setting $J_{ly}+J_{hy}=0$ one obtains the Lorentzian shape longitudinal
magnetoresistance depicted in fig. 2\smallskip 
\begin{eqnarray}
\rho _{xx}(H) \equiv \frac{\varepsilon _{x}}{J_{lx}+J_{hx}}=\rho
_{xx}(H\rightarrow \infty )+\frac{L}{1+\left( H/W\right) ^{2}} \qquad & & \nonumber \\
\rho _{xx}(H \rightarrow \infty )=\frac{%
R_{h}^{2}S_{l}+R_{l}^{2}S_{h}-2QR_{l}R_{h}}{\left[ R_{l}+R_{h}\right] ^{2}} \qquad \qquad \ \ \ \ \ (2) & & \nonumber \\
W=\frac{S_{l}+S_{h}+2Q}{R_{l}+R_{h}};\
L=-\frac{\left[ R_{l}\left( S_{h}+Q\right) -R_{h}\left( S_{l}+Q\right)
\right] ^{2}}{\left( S_{l}+S_{h}+2Q\right) \left( R_{l}+R_{h}\right) ^{2}} & & \nonumber
\end{eqnarray}

The hole densities in the two bands, and hence $R_{l}$, $R_{h}$, are known
from fig. 1. Fitting the data depicted in fig. 2 to eqs. 2, one obtains $\rho
_{xx}(H\rightarrow \infty )$, $L$, and $W$ which are in turn used to extract 
$S_{l}$, $S_{h}$, and $Q$ as a function of temperature and density. The
Landau fan diagram combined with the low field magnetoresistance, hence
provide a unique opportunity to directly measure inter and intraband
scattering. The results of such analysis are shown in fig. 3 where $S_{l}$, $%
S_{h}$, $Q$ , and $\rho _{xx}(H=0)$ are depicted vs. $T$ for the same total
density as in fig. 2. At low temperatures, all these quantities saturate to
some residual values which we attribute to inter and intraband elastic
scattering. As the temperature is increased, inelastic scattering commences
and these quantities grow.

The remarkable and central result of our work is the observation that the
inelastic scattering rates follow the same temperature dependence as $\rho
_{xx}(H=0)$, namely, $S_{i}(T)=S_{i}(0)+\alpha _{S_{i}}\exp (-T_{0}/T)$, $%
Q(T)=Q(0)+\alpha _{Q}\exp (-T_{0}/T)$, where $\alpha _{S_{i}}$, $\alpha _{Q}$
are some constants. Remarkably, the characteristic temperature, $T_{0}$, is
similar to all these quantities, including $\rho _{xx}$! For $S_{l}$ and $%
S_{h}$ we find $T_{0}=5.0K$, for $Q$, $T_{0}=4.8K$, and for $\rho _{xx}$, $%
T_{0}=4.3K$. Similar correlations are found for other total densities in the
"metallic" regime. Our results hence strongly suggest that the universal
Arrhenius temperature dependence of the resistance in the "metallic"
regime merely reflects the increase of inelastic interband scattering with
temperature and probably have nothing to do with phase transition into a
delocalized state.

As expected, the light band is more susceptible to scattering and hence $%
\alpha _{S_{l}}>\alpha _{S_{h}}$. Note the inelastic contributions to $S_{l}$
and $Q$ at $T=2K$ are larger than their elastic counterparts. Moreover, they
are all larger than $\rho _{xx}$, indicating the two bands are strongly
coupled by the scattering. Eventually, for $T>2K$ , the coupling
equilibrates the two drift velocities and the classical magnetoresistance is
fully suppressed. However, the resistance, $\rho _{xx}$, continues to grow.
The latter growth results from inelastic interband scattering that affects
the resistance even when the bands are already fairly strongly coupled. In
fact, in some of the data published in the literature, the Arrhenius
dependence of the resistance is not accompanied by the Lorentzian
magnetoresistance, probably indicating substantial interband scattering.

We find experimentally that $S_{l}(T)\simeq S_{l}(0)+2.1Q;$ $S_{h}(T)\simeq
S_{h}(0)+0.48Q$, thus yielding for the density of figs 2, 3, $\lambda =0.48$.
Since the prefactors multiplying $Q$ are reciprocal, the resistances, $%
S_{l}(0);$ $S_{h}(0),$ are identified as $\sigma _{ll}^{-1}$ and $\sigma
_{hh}^{-1}$, respectively. The diagonal resistances, pertaining to intraband
scattering, are hence found to be practically temperature independent. The
Arrhenius $T$ dependence originates from interband scattering alone.

We turn now to discuss possible reasons for the Arrhenius temperature
dependence of $S_{l},S_{h}$ and $Q$ which in turn lead to a similar
temperature dependence of $\rho _{xx}$. These scattering rates
(expressed as resistances) crucially depend on the bands' dispersions
relations, $E_{i}(\mathbf{k})$, and their resulting excitation
spectra. To extract the bands dispersions depicted in the inset to fig. 4,
we approximate  \cite{Broido1,Ekenberg1} the light band by a parabolic
relation with a mass  \cite{Stormer1} $m_{l}=0.38m_{0}$ ($m_{0}$ is the bare
electron mass). The variation of $p_{l}$, $p_{h}$ with $p_{total}$, depicted
in fig. 1, is then used to calculate the ratio between the two bands
compressibilities. Neglecting band warping as well as differences between
density of states and compressibility, we use the ratio of the two
compressibilities to extract the dispersion of the heavy band. This
dispersion then allows, within the random phase approximation, the
calculation of the excitation spectrum of the system. The spectrum is
composed of two particle-hole continua, one for each band, and two plasmon
branches. The particle-hole continua correspond to regions in the $\mathbf{%
q,\omega }$ plane where the imaginary part of the polarization operator, $%
\Pi (\mathbf{q},\omega )$, is non-zero. The plasmon branches are the poles of
the dielectric constant, $\epsilon (\mathbf{q},\omega )$. Both are shown in
fig. 4 for zero temperature. Due to the very different masses, the optical
plasmon branch corresponds mostly to motion of the light holes while the
acoustic branch originates from the heavy holes screened by the lighter ones
(analogous to acoustic phonons in metals). At small wave-vectors, the
acoustic plasmon velocity is $v_{ap}=\sqrt{\frac{m_{h}}{2m_{l}}}v_{Fh},$
where $v_{Fh}$ is the heavier hole Fermi velocity.

Some interband scattering may be attributed to electron-phonon interaction
but the calculated magnitude of this effect is more than an order of
magnitude too small to account for the measured rates. A more plausible
candidate is interband Coulomb scattering that leads to resistance through
either Coulomb drag or particle transfer.

The rate of interband scattering is proportional to the screened potential
squared, $|\frac{2\pi e^{2}}{q\epsilon (\mathbf{q},\omega )}|^{2}$. In the
absence of plasmons, the dielectric function is regular and at low
temperatures may be approximated by its static value. The resulting rate is
usually proportional to $T^{2}$. The absence of such contribution in our
data is puzzling and calls for a detailed calculation of the scattering
rates with the specific particle-hole continua and band structure depicted
in fig. 4. It may happen that the very different masses, as well as the
concave shape of the heavy band, limit that contribution to small values.

The Coulomb scattering rates may be substantially enhanced in the presence
of plasmons due to the divergent screened interaction in their vicinity,
provided the plasmon branch overlaps the particle-hole continua. This effect
is very pronounced in the calculation of the Coulomb drag between
coupled quantum wells by Flensberg and Hu  \cite{Flensberg1}. As indicated by
fig. 4, at $T=0$ the acoustic plasmon does not intersect the heavy holes
continuum. As the temperature is raised, Im$\left[ \Pi _{h}(\mathbf{q}%
,\omega )\right] $ is thermally activated beyond its zero $T$ boundaries,
leading to a finite overlap with the acoustic plasmon branch. The value of Im%
$\left[ \Pi _{h}(\mathbf{q},\omega )\right]$ is Arrheniusly small 
there but the diverging screened interaction compensates for that.
The corresponding scattering rates depend Arrheniusly on temperature.

We turn now to evaluate the corresponding activation temperature, $T_{0}$.
Since the temperature, $T\leq 2K$ is small compared with the Fermi
energy, we restrict ourselves to small $\mathbf{q}$. A direct calculation
for a concave dispersion relation then gives, Im$\left[ \Pi _{h}(\mathbf{q}%
,v_{ap}q)\right] =\frac{1}{2\sqrt{\pi ^{3}T}}\frac{k_{p}}{\sqrt{\partial
^{2}E/\partial k^{2}}}\exp \left[ -\left( E_{F}-E_{p}\right) /T\right] $.
Here, $k_{p}<k_{Fh}$ is the momentum for which the heavy hole velocity
matches the plasmon velocity, the curvature, $\partial ^{2}E/\partial k^{2}$%
, is evaluated at $k_{p}$ and $E_{p}\equiv E(k_{p})$. Note the result is
independent of $q$! The activation temperature, $T_{0}$, is simply $%
E_{F}-E_{p}$. At our density, $m_{h}=5.15m_{l}$, leading to $v_{ap}\simeq
1.6v_{Fh}$. The corresponding $E_{p}$ is marked in the inset to fig. 4
together with the Fermi energy. The resulting activation energy $T_{0}=E_{F}-E_{p}\simeq 5.5K$, in good agreement with the value
characterizing the inelastic rates, $Q$, $S_{l}$, $S_{h}$ and the
 resistance, $\rho _{xx}$. We are presently calculating the
resulting scattering rates.

We briefly note another possible source of Arrhenius temperature dependence
of the resistance. As the temperature is increased, carriers are transferred
from the light to the heavy band, due to the larger entropy of the latter
(larger density of states). This coexistence of two bands is hence analogous
to the standard liquid (light band) - gas (heavy band) coexistence. Under
certain conditions, the density changes calculated from the corresponding
Clausius-Clapeyron equation may follow an Arrhenius dependence which is
analogous to the vapor pressure in the liquid-gas problem. Since the heavy
hole mobility is lower than that of the light holes, such carrier transfer
should result in resistance increase. We are presently studying the
implications of this effect.

In Summary. We have shown experimentally that the "metallic" behavior of
the resistance of holes in a GaAs/AlGaAs heterostructure results from
inelastic scattering between the two splitted heavy hole bands. The measured
 interband
scattering rates depend Arrheniusly on
temperature with almost the same activation energy as the longitudinal
resistance. Using Shubnikov de Haas data, we mapped the band structure and
the corresponding excitations. We found that due to the dispersion relation
of the heavy band, the system supports a weakly damped acoustic plasmon. The
Arrheniusly small overlap between the heavy hole excitation continuum and
the plasmon branch leads to a similar dependence of the interband scattering
rates, and hence of $\rho _{xx}$ upon $T$. Using the measured band
structures we calculate $T_{0}$ for the plasmon mediated process and obtain
good agreement with the values deduced experimentally from the inelastic
scattering rates. The absence of a substantial power law contribution to the
inelastic interband scattering remains to be explained.

\smallskip

Acknowledgment

This work was supported by the Binational Science Foundation (BSF), Israeli
Academy of Sciences, German-Israeli DIP grant, Minerva foundation, Technion
grant for promotion of research, and by the V. Ehrlich career development
chair.

\medskip

\noindent Figure Captions

\noindent Figure 1 - Hole densities of the two splitted heavy hole bands as a
function of total density. Inset - one of the Shubnikov de Haas traces used
to determine the hole densities.

\smallskip

\noindent Figure 2 - The two-band classical magnetoresistance contribution
to the resistance for different temperatures. Note the perfect Lorentzian
shape. Inset- Full magnetoresistance curves for the same temperatures.

\smallskip

\noindent Figure 3 - The various scattering rates expressed in terms of
resistances (left axis) and the zero field longitudinal resistance (right
axis) vs. $T$. Solid lines depict best fit to Arrhenius dependence.
Inset-same data in $ln$ vs. $1/T$ plot. The slopes give $T_{0}$.

\smallskip

\noindent Figure 4 - Solid lines - heavy and light particle-hole excitation
continua as a function of momentum scaled to the heavy hole Fermi wave
vector. Shaded area corresponds to the range where drag-like interband scattering is possible at very low $T$. Dashed lines -
optical (op) and acoustic (ap) plasmon dispersions. Inset - The measured
bands dispersion relations. The energy $E_{F}$ corresponds to the 
hole Fermi energy and $E_{p}$ is the energy where the heavy hole velocity
matches that of the acoustic plasmon. The difference, $E_{F}-E_{p}$ gives
the activation temperature, $T_{0}$ (see text).

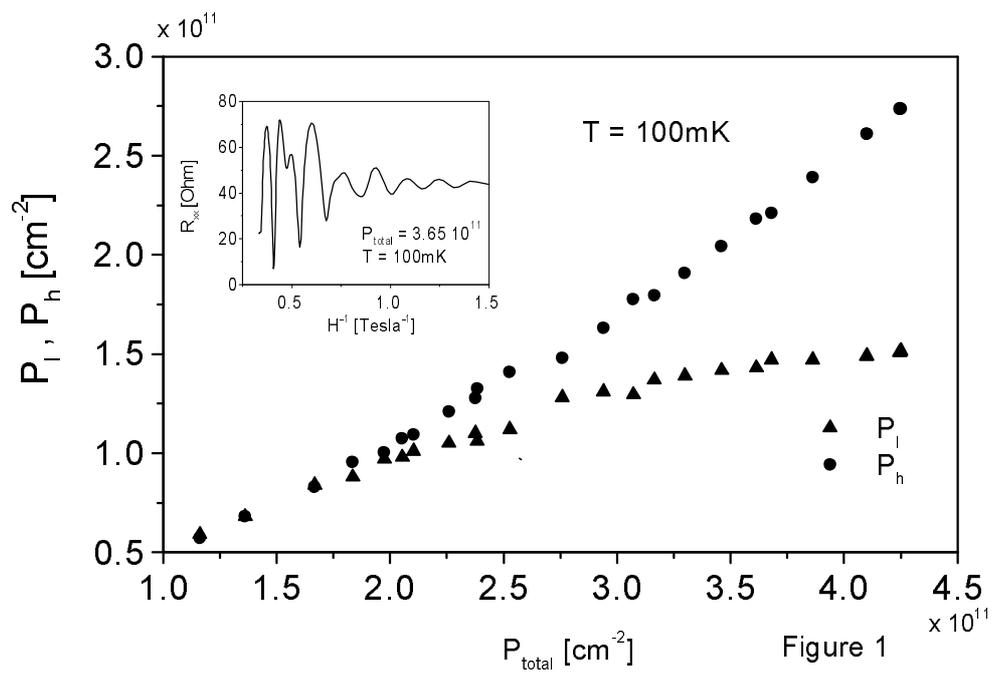
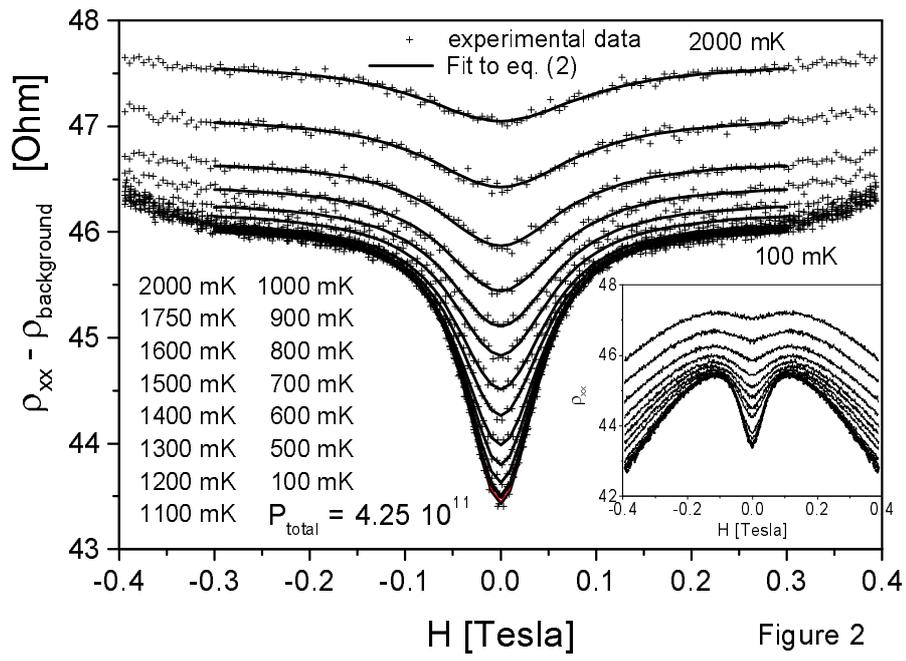
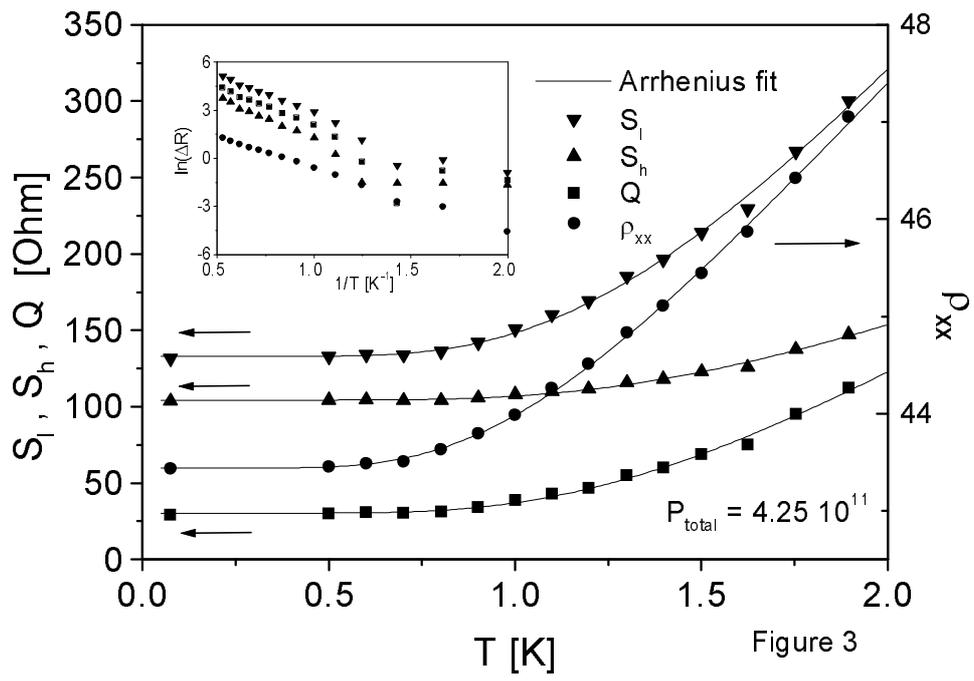
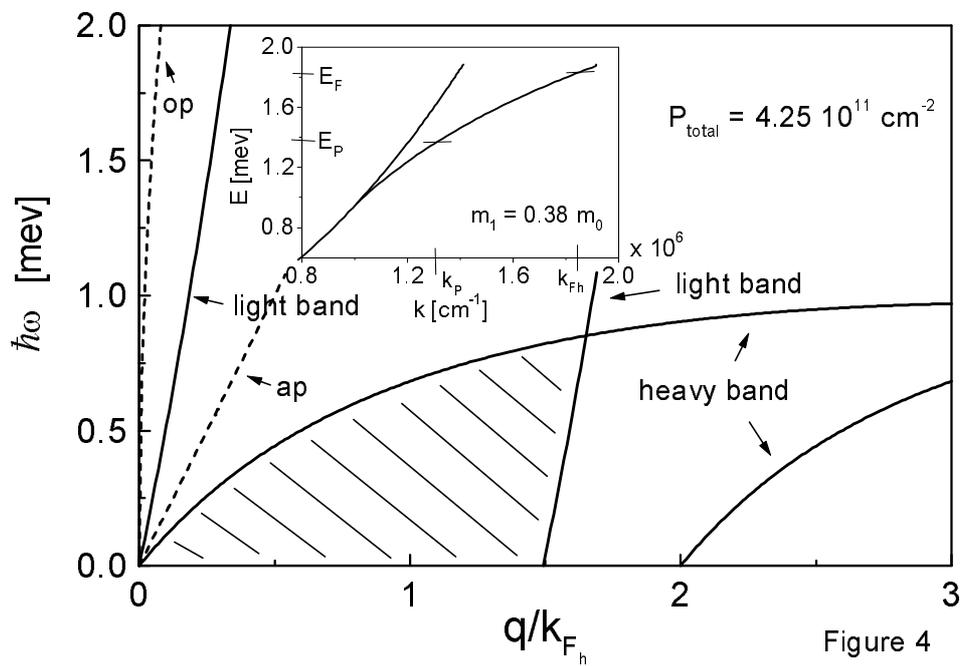
\end{document}